# Comparison of near-field light intensities: plasmon nanofocusing vs localized plasmon resonance

Tongyao Li[1], Andrea Schirato[2,3], Taku Suwabe[1], Remo Proietti Zaccaria[4], Prabhat Verma[1], and Takayuki Umakoshi[1,5]*

1. Department of Applied Physics, Osaka University, 2-1 Yamadaoka, Suita, Osaka, 565-0871 Japan.
2. Department of Physics, Politecnico di Milano, Piazza Leonardo da Vinci 32, Milano, 20133 Italy.
3. Department of Physics and Astronomy, Rice University, Houston, Texas 77005, United States
4. Istituto Italiano di Tecnologia, Via Morego 30, Genova, 16163 Italy
5. Institute for Advanced Co-Creation Studies, Osaka University, 2-1 Yamadaoka, Suita, Osaka, 565-0871 Japan.

*E-mail: umakoshi@ap.eng.osaka-u.ac.jp

**Abstract**

The localized surface plasmon resonance of metallic nanostructures produces strongly localized and enhanced near-field light, significantly contributing to nanophotonics research and applications. Plasmon nanofocusing represents another method for generating near-field light through the propagation and condensation of plasmons on tapered plasmonic structures. In both methods, the intensity of near-field light is a critical aspect for many applications. In this study, we numerically inspect and compare the intensities of near-field light generated by either localized plasmon resonance or plasmon nanofocusing. To account for the light-induced changes in the optical properties of plasmonic structures, which in turn influence the near-field light intensity, we couple electromagnetic and thermal calculations to consider in a fully self-consistent manner the effects of the incident light and the light-induced temperature rise within the metal. A gold nanorod and a cone were adopted for exciting the localized plasmon resonance and plasmon nanofocusing, respectively. We find that plasmon nanofocusing generates approximately 1.5 times as strong near-field light as localized plasmon resonance. Our research provides a necessary foundation for generating near-field light, which is crucial for advancing the applications of near-field optics.

Strongly localized and enhanced near-field light produced by plasmonic nanostructures has been widely applied in various fields from material science to biology, having realized a number of optical applications in the past few decades, such as enhanced optical spectroscopy, super-resolution microscopy, light harvesting, and plasmonic lasers. [1-6]. Near-field light is usually achieved by exciting the localized surface plasmon resonance (LSPR) of metallic nanostructures that work as optical antennas, with gold nanoparticles and nanorods being among the most typically employed systems [7-8]. In recent years, plasmon nanofocusing has caught considerable attention as another method for generating near-field light. In the process of plasmon nanofocusing, surface plasmon polaritons propagate along a metallic tapered structure, such as a gold conical structure, toward the apex while compressing their energy, and eventually create strong near-field light at the nanometrically sharp apex [9-12]. A grating structure fabricated on the shaft of a tapered structure is often used as a coupler to excite plasmons. Because the grating structure is located far from the apex, one of the advantages of nanofocusing is that it is background-free from incident light, contrary to the LSPR approach, where the direct incident light illumination of plasmonic nanostructures spatially overlaps with near-field light [13-16]. The broadband characteristic of plasmon nanofocusing has also recently been recognized as another unique property. Plasmon nanofocusing can be achieved over a wide frequency range, because it is based on the propagation of plasmons, unlike plasmon resonances occurring within a specific spectral range [17-20].

Considering the two methods of LSPR and plasmon nanofocusing, one of the fundamental but pivotal questions is which method generates more intense near-field light, as the near-field light intensity is a basic and important optical property for many applications. With respect to LSPR, the strength of near-field light has often been described as the enhancement factor, which is calculated as the ratio of the near-field light intensity to the incident light intensity. For plasmon nanofocusing, the amount of incident light energy converted to near-field light, that is, the conversion efficiency, has usually been investigated. Although the enhancement factor and efficiency are important for evaluating the near-field light intensity, they are normalized by the incident light intensity. As such, they do not provide direct information on the absolute value of the near-field light intensity for both methods, which in fact can also be critical from an application standpoint. In particular, upon increasing the incident light intensity, which in turn increases the near-field light intensity, plasmonic structures can be melted and irreversibly destroyed due to the heat generated by ohmic losses [21-25]. This suggests that the maximum threshold incident light intensity can be defined as the one causing a temperature increase right below the melting point of plasmonic structures. In these terms, the photogeneration of heat contributing to the temperature rise is a key factor to gauge the maximum allowed incident light intensity and the corresponding maximum near-field light intensity for both LSPR and plasmon nanofocusing.

In this study, we numerically investigated which between LSPR and plasmon nanofocusing generates more intense near-field light, by including the effects of light-induced heat generation and temperature rise in plasmonic structures. We selected a gold nanorod and a gold cone as the plasmonic systems typically used for the excitation of LSPR and plasmon nanofocusing, respectively. We examined the temperature and near-field light intensity by varying the incident light intensity. The maximum incident light intensities and the resulting maximum near-field light intensities were evaluated, and were found to be different between LSPR and plasmon nanofocusing, owing to the difference in the geometrical configurations between the gold nanorod and cone. We found that plasmon nanofocusing allows for achieving a higher near-field light intensity than LSPR when optimal input powers were exerted on each structure. As for plasmon nanofocusing, we also observed a saturation followed by an inverse change in the near-field light intensity with respect to the incident light intensity, which our model explained by coupling electromagnetic and thermal problems consistently. Finally, since the melting temperature of gold (1337 K) is relatively high for most applications, we also considered moderate illumination conditions and temperature increases up to 40°C (313.15 K) and 100°C (373.15 K), which are relevant for practical situations [26].

We used the finite element method-based commercial software COMSOL Multiphysics to calculate the steady-state electric field intensity and temperature field of the plasmonic structures under scrutiny. The schematics of our numerical models are shown in Fig. 1. We chose 785 nm as the incident light wavelength, which is typical in various optical measurements and applications. The incident light has a Gaussian space profile, with a beam waist of 550 nm. Both nanostructures are embedded in air. For the gold nanorod, we set the rod diameter and length to 20 and 105 nm, respectively, to tune the longitudinal LSPR wavelength to 785 nm (Fig. S1). As for the gold cone, the cone apex size is 20 nm, which is the same as the gold nanorod diameter, so that the plasmonic confinement of near-field light is comparable with that of the nanorod. The cone length was set to 7 μm. We designed a grating structure 3.75 μm away from the apex, which is far enough to separate the incident light from the near-field light. It is composed of three grooves, whose period and depth are 680 and 100 nm, respectively, and it is optimized for the best plasmon coupling efficiency at a wavelength of 785 nm, as shown in Fig. S2. The entire gold nanorod was exposed to illumination, whereas the grating structure was irradiated by the incident light for the gold cone, as illustrated in Figs. 1(a) and 1(b). Therefore, the grating can receive a greater number of photons owing to its large structure compared with the small nanorod in the same illumination condition, which is another characteristic of plasmon nanofocusing.

For the gold nanorod and the cone, the electric field intensity and temperature were calculated by solving Maxwell's equations coupled to the heat transfer. The initial temperature was set to room temperature (293.15 K/20 °C). When the incident light is irradiated, not only near-field light but also Joule heating is generated, following photon absorption and subsequent electromagnetic dissipation, which leads to a temperature increase in these metallic structures [27, 28]. In turn, the permittivity of gold is modified by such an increase of the metal temperature. Specifically, we accounted for this effect by considering a temperature-dependent damping factor in an analytical Drude-Lorentz-like permittivity, which is suited for gold in this wavelength range [29-31]. The change in the permittivity of gold alters the electromagnetic response of the plasmonic system, including the near-field light intensity and electric field inside the gold structures, causing a change in the amount of Joule heating. Therefore, by coupling the optical and photothermal responses of the considered nanostructures, our numerical model treats this interdependence fully consistently until convergence to the steady state. The calculation details are thoroughly described in the Supporting Information Note S1.

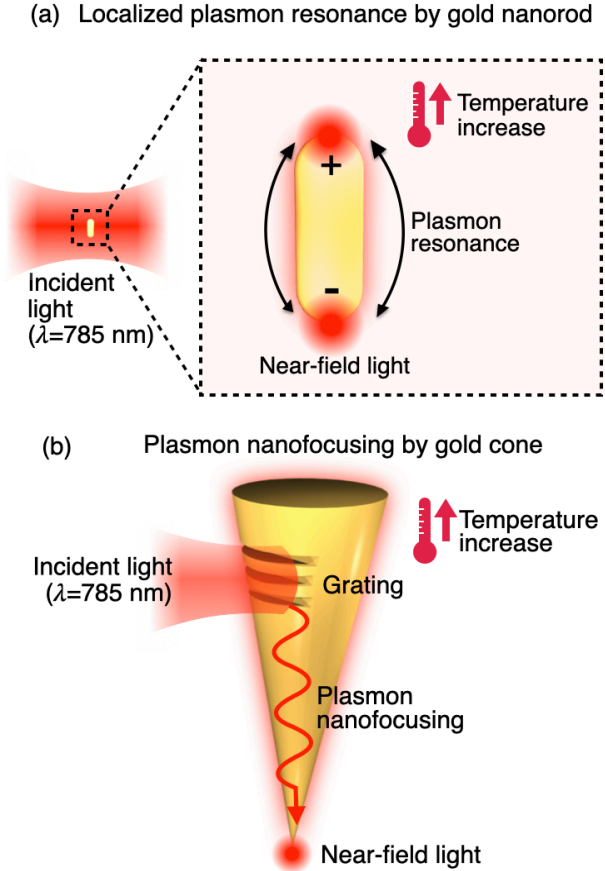

**Figure 1.** Schematics of calculation models of (a) localized plasmon resonance and (b) plasmon nanofocusing

We first investigated the effects of exciting the LSPR of the nanorod. As shown in Fig. 2(a), when the gold nanorod is irradiated, near-field light appears at both ends of the nanorod, which is the typical electric field distribution of the longitudinal plasmonic mode of nanorods. Here, the incident light intensity was set to 1.33 × $10^7$ W/m², by referring to the maximum intensity at the center of the Gaussian distribution of the incident light, which corresponds to a total power of 12.6 μW. This incident intensity produced a near-field light intensity of 2.75 × $10^9$ W/m², indicating an approximately 207 times intensity enhancement. In particular, the near-field light intensity was monitored 3 nm away from the end of the gold nanorod. As expected, when the incident light intensity increased, the near-field light intensity also increased, as shown in Figs. 2(b) and 2(c). Specifically, we obtained near-field light intensities of 14.92 × $10^9$ and 47.77 × $10^9$ W/m² when the incident light intensities increased to 8.31 × $10^7$ and 47.79 × $10^7$ W/m², indicating intensity enhancements of ~ 180 and ~ 101 times, respectively. Interestingly, the enhancement of the electric field intensity decreased with increasing incident light intensity. We ascribe this saturation effect to the larger damping factor at higher temperatures, as described in detail in the following. For comparison, results obtained assuming a constant damping factor are also presented (see Fig. S3). The temperature distributions were then calculated for the same incident light intensities, as shown in Figs. 2(d)-2(f). The temperature was uniform over the gold nanorod because of the high thermal diffusion of the metal, in stark contrast with the inhomogeneous electric field distributions. As expected, the temperature also increased as the incident light intensity increased. When the incident light intensity was $1.33 \times 10^7$ W/m², the temperature, evaluated as the surface average over the entire gold nanorod, increased from 293.15 K to 318 K. It increased to 418 and 721 K with incident light intensities of $8.31 \times 10^7$ and $47.79 \times 10^7$ W/m², respectively.

To further inspect the dependence of the near-field light intensity and light-induced temperature increase upon increasing the illumination intensity, as shown in Fig. 2(g), we varied the incident light intensity up to $2.80 \times 10^9$ W/m², to reach a temperature close to the melting point of gold (1337 K, as indicated by the red dashed line). Both the near-field light intensity and temperature increased as the incident light intensity increased, yet following a sub-linear behavior. The nonlinear change in the near-field light intensity is explained as the result of the change in the damping factor of gold. We found that the maximum near-field light intensity, that is, slightly below the melting point, was $112.70 \times 10^9$ W/m², corresponding to an incident light intensity of $2.39 \times 10^9$ W/m².

Please note that the melting point of 1337 K is for bulk gold [32]. However, the melting point can be reduced for nano-sized materials due to the effects of size. In fact, the reduction in the melting point was estimated to be only a few tens of degrees for a gold nanosphere with a diameter of 10 nm, according to previous studies [32-34]. Compared to the 10-nm-large gold nanospheres, the size effect should be much smaller for the 105-nm-long gold nanorod. In addition, it is not straightforward to accurately evaluate the size effect on the melting point decrement of the gold nanorod. Therefore, we ignored the size effect and used the melting point of bulk gold in this study. If the size effect is considered, the maximum incident light intensity and the resulting near-field light intensity should be slightly lower.

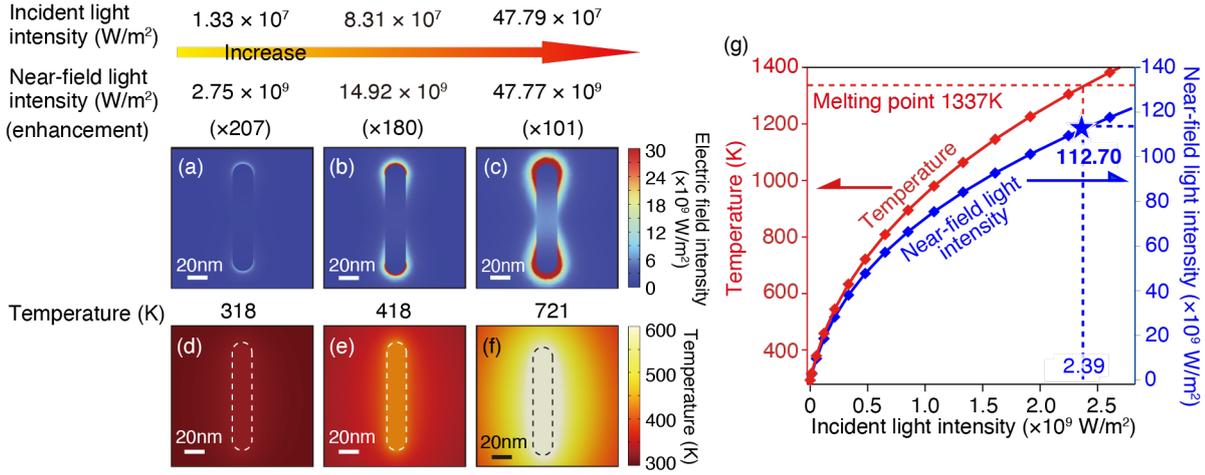

**Figure 2.** Electric field distributions at incident light intensities of (a) $1.33 \times 10^7$, (b) $8.31 \times 10^7$, and (c) $47.79 \times 10^7$ W/m². Temperature distributions at the same incident light intensities of (d) $1.33 \times 10^7$ (318 K), (e) $8.31 \times 10^7$ (418 K), and (f) $47.79 \times 10^7$ (721 K). (g) Dependence of near-field light intensity and temperature of the gold nanorod on the incident light intensity.

For plasmon nanofocusing on the gold cone, we also investigated the electric field intensity and temperature in a manner similar to that detailed above for the gold nanorod. Figure 3(a) shows the electric field distribution map around the gold cone structure under the incident light illumination of the grating coupler with an intensity of $107.73 \times 10^7$ W/m². Surface plasmons were excited at the grating and propagated along the surface of the gold cone structure, generating highly confined near-field light at the apex. The enlarged map around the apex is shown in Fig. 3(d), which clearly confirms plasmon nanofocusing inducing the near-field light at the apex. The near-field light intensity obtained at 3 nm away from the tip apex was $94.09 \times 10^9$ W/m², indicating an intensity enhancement of ~ 87 times. The temperature distribution under the same condition is shown in Fig. 3(b). Similar to the gold nanorod, the temperature uniformly increased to 420 K over the entire gold cone. The enlarged temperature map around the apex is also shown in Fig. 3(g). We then reduced the incident intensity to $11.97 \times 10^7$ W/m² (Fig. 3(c)) and increased it to $430.92 \times 10^7$ W/m² (Fig. 3(e)). With a smaller incident light intensity of $11.97 \times 10^7$ W/m², we obtained a near-field light intensity of $13.08 \times 10^9$ W/m². The near-field light intensity was $178.39 \times 10^9$ W/m² in the case of a larger incident light intensity of $430.92 \times 10^7$ W/m². Therefore, similar to the case of LSPR, the near-field light intensity increased with the incident light intensity in the plasmon nanofocusing. Accordingly, the temperature was also increased from 307 to 805 K by increasing the incident light intensity from $11.97 \times 10^7$ to $430.92 \times 10^7$ W/m², as shown in Figs. 3(f) – (h).

Furthermore, we extensively investigated the dependences of near-field light intensity and temperature increase on the incident light intensity, as shown in Fig. 3(i). We changed the incident light intensity up to $10.00 \times 10^9$ W/m². Interestingly, the near-field light intensity first increases with increasing incident light intensity, and subsequently decreases even though the incident light intensity increases. In contrast, the temperature monotonically increases with the incident light intensity across the entire range we spanned. The linear increase in temperature, which differs from the nonlinear behavior of the gold nanorod, can be due to the size of the gold cone. The gold cone is much larger than the gold nanorod, which can lead to a larger heat capacity and a linear temperature increase. The counterintuitive highly nonlinear inverse trend in the near-field light intensity is explained by the fact that, as mentioned previously, a higher temperature produces a larger damping factor of gold, which negatively influences the near-field light intensity to be reduced. Although increasing the incident light intensity simply increases the near-field light intensity, the higher temperature caused by the larger incident light intensity causes a reduction in the near-field light intensity. Therefore, depending on the balance between these two effects, the near-field light intensity can be decreased even upon increasing the incident light intensity. In particular, because plasmon nanofocusing involves the propagation of plasmons for a certain distance, the damping factor can dominate the process to a larger extent than LSPR, which does not involve propagation. We therefore evaluated the decrement of near-field light intensity due to larger damping factors at high temperatures for plasmon nanofocusing. We found indeed that this effect impacted more dominantly than the increment of near-field light intensity for increasing incident light intensities, leading thus to the reduction of near-field light intensity (as discussed in detail in Supporting Information Note S5).

In plasmon nanofocusing, we found that the maximum near-field light intensity of $178.45 \times 10^9$ W/m² was achieved for an incident light intensity of $4.55 \times 10^9$ W/m², where the temperature was 834 K. The near-field light intensity monotonically decreased for incident light intensities greater than this value. When the temperature approached the melting point, the near-field light intensity was $138.53 \times 10^9$ W/m², where the incident light intensity was $8.99 \times 10^9$ W/m². Therefore, we concluded that the maximum near-field light intensity obtained by plasmon nanofocusing on the gold cone was $178.45 \times 10^9$ W/m². As the maximum near-field light intensity was 112.70 × 10$^9$ W/m² for LSPR, plasmon nanofocusing induces near-field light approximately 1.5 times stronger than that for LSPR.

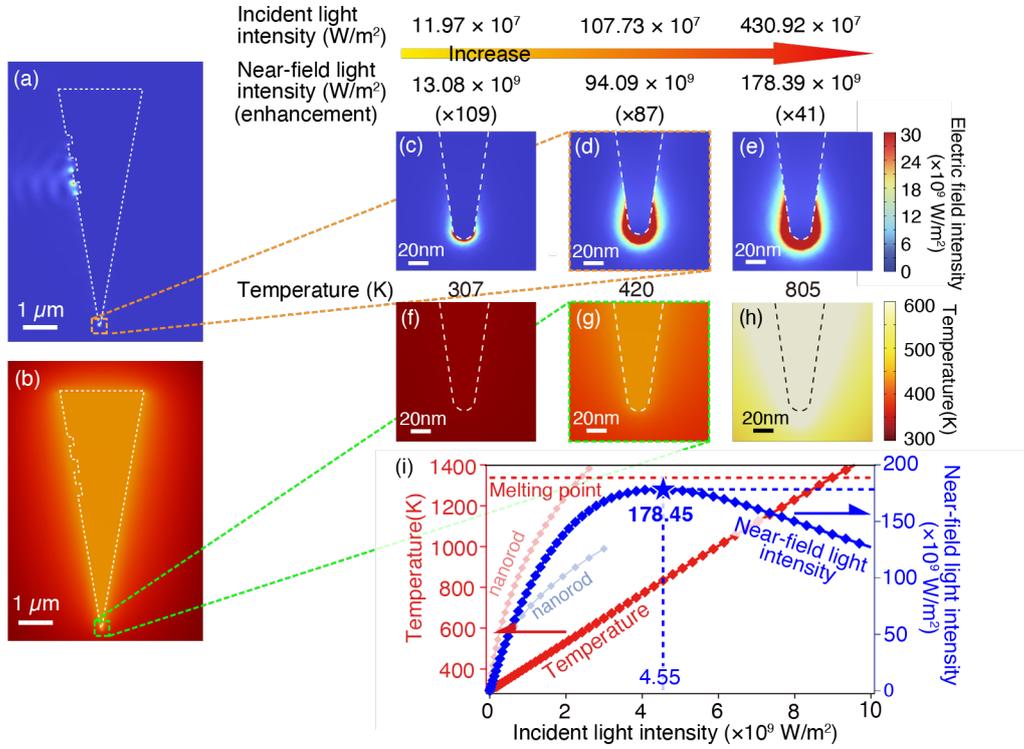

**Figure 3.** (a) Electric field distribution and (b) temperature distribution of the gold cone at an incident light intensity of 107.73 × 10$^7$ W/m². Electric field distribution maps enlarged near the apex at incident light intensities of (c) 11.97 × 10$^7$, (d) 107.73 × 10$^7$, and (e) 430.92 × 10$^7$ W/m². Temperature distribution maps enlarged near the apex at incident light intensities of (f) 11.97 × 10$^7$ (307 K), (g) 107.73 × 10$^7$ (420 K), and (h) 430.92 × 10$^7$ W/m² (805 K). (i) Dependence of the near-field light intensity and temperature of the gold cone on the incident light intensity. For comparison, the near-field light intensity and temperature obtained for the gold nanorod are also shown as light blue and red curves, respectively, which were obtained from Fig. 2(g).

So far, we have considered a temperature close to the melting point; however, this is not practical as the temperature of the melting point is too high for most applications and rather sets an upper boundary of operation. Therefore, we investigated the illumination conditions required to reach moderate maximum temperatures, such as 40°C (Fig. 4(a)) and 100°C (Fig. 4(b)). Here, we increased the incident light intensity up to 80.00 × 10$^7$ W/m². Temperatures of 40°C (313.15 K) and 100°C (373.15 K) are indicated by the green and orange dashed lines, respectively. Both the near-field light intensities and temperatures increase as the incident light intensity increases for both the gold nanorod and cone. It is evident that the temperature of the gold nanorod rises quickly with the incident light intensity due to its smaller size compared to the gold cone. Therefore, the temperature easily reached the limit (40 or 100°C) with low incident light intensities. At a temperature of 40°C, the near-field light intensity generated by LSPR was only $2.41 \times 10^9$ W/m$^2$ (corresponding to an incident light intensity of $1.08 \times 10^7$ W/m$^2$), whereas the near-field light intensity generated by plasmon nanofocusing was $18.35 \times 10^9$ W/m$^2$ (incident light intensity: $16.99 \times 10^7$ W/m$^2$). We thus found that plasmon nanofocusing generated near-field light ~7.6 times stronger than that generated by LSPR to reach the same temperature. Considering a temperature of 100°C, the near-field light intensity was $9.63 \times 10^9$ W/m$^2$ for LSPR (incident light intensity of $4.90 \times 10^7$ W/m$^2$). In contrast, the near-field light intensity generated by plasmon nanofocusing was $64.87 \times 10^9$ W/m$^2$ (incident light intensity: $67.77 \times 10^7$ W/m$^2$), which was ~6.7 times stronger than that of LSPR. These results indicate that plasmon nanofocusing is more advantageous as it generates significantly stronger near-field light compared to LSPR when used in a practical situation at a moderate temperature.

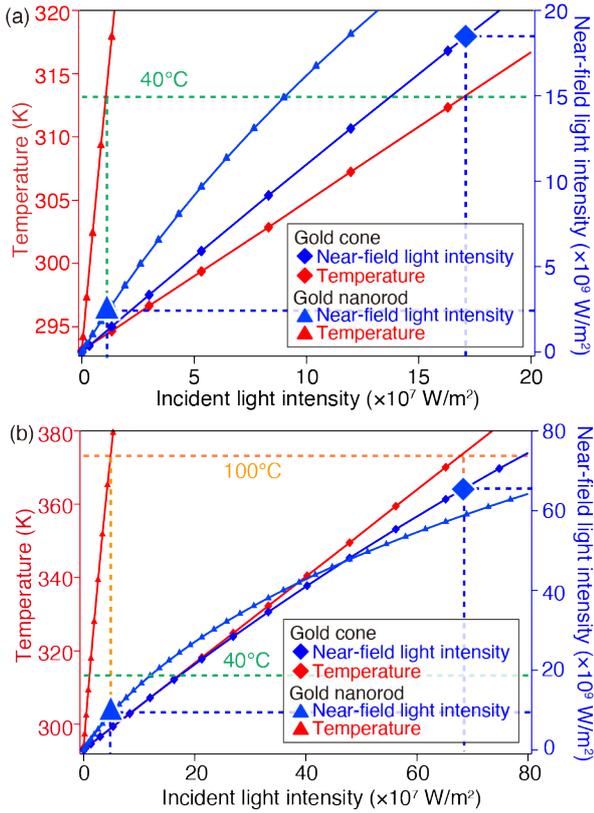

**Figure 4.** Dependences of the near-field light intensity and temperature on the incident light intensity at a lower intensity range for the gold nanorod and cone. Lower temperature limitations of (a) 40°C and (b) 100°C were considered.

In conclusion, we investigated which between LSPR or plasmon nanofocusing generates stronger near-field light. The maximum near-field light intensity was determined by considering the light-induced heat generation and temperature increase in the gold structures. It was $112.70 \times 10^9$ W/m² for LSPR of the gold nanorod, whereas it was $178.45 \times 10^9$ W/m² for plasmon nanofocusing of the gold cone, when the incident light intensities were $2.39 \times 10^9$ and $4.55 \times 10^9$ W/m², respectively. Therefore, plasmon nanofocusing generates approximately 1.5 times as strong near-field light as that generated by LSPR. The input powers are in the range of several milliwatts within a micrometer-order focal spot, which is not far from our expectation of causing the destruction of gold nanostructures [35]. We thus believe that our simulation results are within a reasonable range compared with actual experiments. Note that some differences are anticipated between our simulations and actual experiments. For example, the gold structures suspended in air were considered in our simulations, whereas gold structures are usually placed on a substrate in actual experiments, which can have a relevant role in dissipating heat. Considering lower temperature regions, plasmon nanofocusing was more effective, as several times stronger near-field light was obtained compared with LSPR. In this study, we only considered the specific condition with the gold nanorod and cone with an incident wavelength of 785 nm. The results can differ between different materials and shapes of the plasmonic structures. More extensive studies as future works under various conditions, exploiting the comparative approach proposed here, are crucial for elucidating the near-field light intensity for both LSPR and plasmon nanofocusing. This fundamental study provides important and practical insights relevant to various applications based on nanophotonics and plasmonics.

**Acknowledgment**

T.U. acknowledges the JSPS KAKENHI Grant-in-Aid for Scientific Research (B) (No. 24K01385), Grant-in-Aid for Transformative Area (A) "Evolution of Chiral Materials Science using Helical Light Fields" (No. 23H04590), Grant-in-Aid for Transformative Area (A) "Material Science of Meso-Hierarchy" (No. 24H01717), Grant-in-Aid for Challenging Research (Exploratory) (No. 24K21718), JST FOREST (No. JPMJFR233Z), and the Takahashi Industrial and Economic Research Foundation. A.S. acknowledges the European Union's Horizon Europe Research and Innovation Program under the Marie Skłodowska-Curie Action PATHWAYS HORIZON-MSCA-2023-PF-GF grant agreement No. 101153856.

**References**

1. S. Kawata, et al., Nat. Photon. 3, 388–394 (2009)
2. P. Verma, Chem. Rev. 117, 6447–6466 (2017)
3. T. Umakoshi, et al., Sci. Rep. 12, 12776 (2022)
4. Y. Tian, et al., J. Am. Chem. Soc. 127, 7632–7637 (2005)
5. R. F. Oulton, et al., Nature 461, 629–632 (2009)
6. Y. J. Lu, et al., Science 337, 450–453(2012)
7. W. L. Barnes, et al., Nature 424, 824–830 (2003)
8. L. Vigderman, et al., Adv. Mater. 24, 4811–4841 (2012)
9. A. J. Babadjanyan, et al., J. Appl. Phys. 87, 3785–3788 (2000)
10. M. I. Stockman, et al., Phys. Rev. Lett. 93, 137404 (2004)
11. H. Choo, et al., Nat. Photon. 6, 838–844 (2012)
12. R. P. Zaccaria, et al., Phys. Rev. B 86, 035410 (2012)
13. C. Ropers, et al., Nano Lett. 7, 2784–2788 (2007)
14. T. Umakoshi, et al., Nanoscale. 8, 5634–5640 (2016）
15. C. C. Neacsu, et al., Nano Lett. 10, 592–596 (2010)
16. S. Berweger, et al., J. Phys. Chem. Lett. 3, 945–952 (2012)
17. T. Umakoshi, et al., Sci. Adv. 6, eaba4197 (2020)
18. K. Taguchi, et al., J. Phys. Chem. C. 125, 6378–6386 (2021)
19. M. Esmann, et al., Nat. Nanotechnol. 14, 698–704 (2019)
20. V. Kravtsov, et al., Nat. Nanotechnol. 11, 459–464 (2016)
21. X. M. Cui, et al., Chem. Rev. 123, 6891–6952 (2023)
22. L. Jauffred, et al., Chem. Rev. 119, 8087–8130 (2019)
23. B. Yang, et al., Adv. Mater. 34, 2107351 (2022)
24. A. Lalisse, et al., J. Phys. Chem. C. 119, 25518–25528 (2015)
25. K. Nishita, et al., ACS Photon. 7, 2139–2146 (2020)
26. G. Baffou, et al., Nat. Mater. 19, 946-958 (2020)
27. A. O. Govorov, et al., Nano Today 2, 30-38 (2007)

28. G. Baffou, et al., Laser Photonics Rev. 7, 171-187 (2012)
29. A. D. Rakic, et al., Appl. Opt. 22, 5271–5283 (1998)
30. P. B. Johnson, et, al., Phys. Rev. B 6, 4370–4379 (1972)
31. A. Alabastri, et, al., Materials 6, 4879–4910 (2013)
32. Ph. Buffat, et al., Phys. Rev. A 13, 2287–2298 (1976)
33. G. L. Allen, et al., Thin Solid Films 144, 297–308 (1986)
34. A. Safaei, et, al., J. Phys. Condens. Matter. 19, 216 (2007)
35. M. Honda, et al., Opt. Express 19, 12375–12383 (2011)

*Supplemental materials for*

# Comparison of near-field light intensities: plasmon nanofocusing vs localized plasmon resonance

Tongyao Li[1], Andrea Schirato[2,3], Taku Suwabe[1], Remo Proietti Zaccaria[4], Prabhat Verma[1], and Takayuki Umakoshi[1,5]*

1.  Department of Applied Physics, Osaka University, 2-1 Yamadaoka, Suita, Osaka, 565-0871 Japan.
2.  Department of Physics, Politecnico di Milano, Piazza Leonardo da Vinci 32, Milano, 20133 Italy.
3.  Department of Physics and Astronomy, Rice University, Houston, Texas 77005, United States.
4.  Istituto Italiano di Tecnologia, Via Morego 30, Genova, 16163 Italy.
5.  Institute for Advanced Co-Creation Studies, Osaka University, 2-1 Yamadaoka, Suita, Osaka, 565-0871 Japan.

*E-mail: umakoshi@ap.eng.osaka-u.ac.jp

# Note S1. Details of the numerical calculations

We used the commercial software COMSOL Multiphysics, version 6.1, based on the finite element method (FEM) to calculate the steady-state electric field intensity and temperature fields for the localized surface plasmon resonance excitation of the gold nanorod, and for the plasmon nanofocusing effect in the gold cone. In our simulations, the incident light had a Gaussian distribution with a beam waist of 550 nm and a wavelength of 785 nm. By setting the incident electric field that describes such excitation conditions as the input background field, the fully-vectorial wave equation that we solved numerically to determine the distribution of the electric field $E$, solution of the electromagnetic problem in the scattering formalism across the computational domains, reads as follows:

$$\nabla \times (\nabla \times \mathrm{E}) - k_0^2 \varepsilon E = 0.$$

Here, $k_0$ is the light wave-vector in free space. The relative permittivity $\varepsilon$ in the surrounding environment was set to 1 (air), while in the gold domain it was defined based on the Drude-Lorentz model, which is expressed as

$$\varepsilon = \varepsilon_\infty + \sum_{j=1}^{M} \frac{f_j \omega_p^2}{\omega_{0j}^2 - \omega^2 + i\Gamma(T)\omega},$$

where $\varepsilon_\infty$ is the permittivity in the high-frequency limit, $\omega_p$ is the plasma frequency of gold, $\omega_{0j}$ is the resonant frequency of the $j$-th Lorentz oscillator, $\omega$ is the angular frequency of the electric field, $\Gamma(T)$ is the damping factor, which is dependent on the temperature $T$, and $f_j$ is the $j$-th oscillator strength. In this study, we considered up to 5 oscillators.

Based on the calculated $E$, the light-induced heat generation within the metal and the subsequent stationary temperature increase were determined. The Joule heating dissipation term $Q$ is related to the electric field intensity $E$ according to the following expression:

$$Q = \frac{1}{2} \mathrm{Re}\{E \cdot j_\mathrm{D}^*\}.$$

Here, $j_\mathrm{D} = i\omega\varepsilon_0\varepsilon E$ is the induced displacement current density, $\varepsilon_0$ is the vacuum permittivity, $\cdot^*$ denotes the complex conjugate, and $\mathrm{Re}\{\cdot\}$ is the real part. The initial temperature was set to room temperature (293.15 K). The temperature variation caused by electromagnetic dissipation within the gold domain was calculated using the heat diffusion equation, reading as follows:

$$\nabla \cdot (-\kappa \nabla \mathrm{T}) = Q,$$

where κ is the thermal conductivity of the considered material [either air. 0.0257 W/(m·K), or gold, 318 W/(m·K)], and $Q$ is the calculated Joule heat. The left-hand side of the equation describes the spatial diffusion of heat via thermal conduction, and the right-hand side is the heat source term. Additional heat exchange terms, such as surface radiation and convection at the surfaces of the gold nanostructures, have been disregarded, since they typically contribute to a negligible extent to the thermal transport in nanosystems like the ones under consideration.

Based on the calculated temperature, the temperature-dependent damping factor $\Gamma(T)$ changes, which therefore modifies the permittivity of gold $\varepsilon$. The damping factor $\Gamma(T)$ is divided into three components, as follows:

$$\Gamma = \Gamma_{\mathrm{e-e}} + \Gamma_{\mathrm{e-ph}} + \Gamma_{\mathrm{surf}}.$$

$\Gamma_{\mathrm{e-e}}$ expresses the damping between electrons, $\Gamma_{\mathrm{e-ph}}$ describes the damping factor between electrons and phonons, and $\Gamma_{\mathrm{surf}}$ represents the damping factor between electrons and metal surfaces. They are expressed as:

$$\Gamma_{e-e} = \frac{\pi^3 \phi \Delta}{12 h E_f}\{(k_B T)^2 + (\hbar\omega)^2\},$$

$$\Gamma_{e-ph} = \Gamma_0 \left(\frac{5}{2} + \frac{4T^5}{\theta_D^5}\int_0^{\frac{\theta_D}{T}} \frac{z^2}{e^z - 1} dz\right),$$

$$\Gamma_{surf} = \frac{S}{4V} A v_f.$$

The coefficients are summarized in the following table. Each value was taken from a previous study [1]. According to the calculated temperature $T$, each damping factor component and total damping factor $\Gamma(T)$ were calculated.

Table S1. The coefficients for each damping factor component

| | |
|---|---|
| $\phi$ | Fermi-surface average of scattering probability |
| $\Delta$ | fractional Umklapp scattering coefficient |
| $h$ | Plank constant |
| $E_f$ | Fermi energy |
| $k_B$ | Boltzmann constant |
| $\Gamma_0$ | bulk damping constant |
| $\theta_D$ | Debye temperature |
| $S$ | surface area of the substance |
| $V$ | volume of the substance |
| $A$ | coefficient considering the scattering mechanism |
| $v_f$ | Fermi velocity |

A change in the damping factor modifies the permittivity and electric field, which in turn modifies the heat generation and temperature. As these properties are correlated with each other, a stationary segregated solver that treats them as coupled is defined in the model.

# Note S2. Calculation model for the gold nanorod

The length and diameter of the gold nanorod were designed to be 105 and 20 nm, respectively, as shown in Fig. S1(a), such that its longitudinal plasmonic resonance wavelength matches the incident wavelength of 785 nm, as shown in Fig. S1(b). The polarization of the incident light is aligned with the longitudinal axis of the rod. In the simulations, the nanostructure is embedded in a homogeneous environment (made of air), defining a sphere of 500 nm radius, *i.e.*, large enough compared to the nanorod size. The computational domain is further surrounded by a fictive spherical domain (400-nm thickness), used to define respectively a perfectly matched layer (PML) for the electromagnetic calculations, and an infinite elemental domain (IED) for the temperature ones, as shown in Fig. S1(a). Beyond the PMLs, scattering boundary conditions are set, while beyond the IEDs, a fixed temperature equal to room temperature is enforced. As described above, the incident light had a Gaussian distribution with a beam radius of 550 nm, which was large enough to cover the nanorod structures. The gold nanorod was placed at the center of the incident beam.

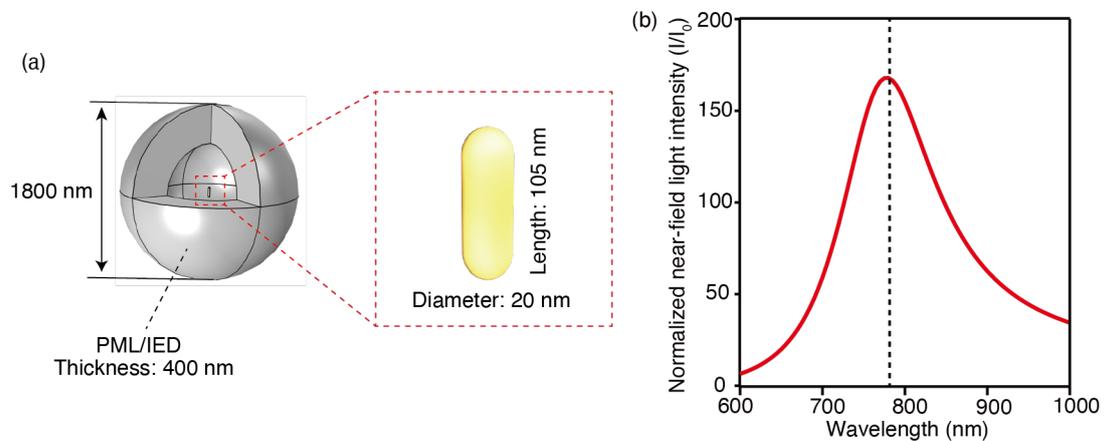

**Figure S1.** (a) Schematic of the calculation model for the gold nanorod. (b) Near-field light intensity spectrum of the gold nanorod normalized by incident light intensity.

# Note S3. Calculation model for the gold cone

For plasmon nanofocusing of the gold cone, the cone length was 7 μm, and the grating was placed 3.75 μm away from the cone apex, as shown in Fig. S2(a). The diameter of the gold cone apex was 20 nm, which is the same as the diameter of the gold nanorod for comparable field confinement. The cone angle was 28 degrees. The grating period and depth were 680 and 100 nm, respectively, which realized optimal coupling with an incident light wavelength of 785 nm, as shown in Fig. 2(b). Incident light with a beam radius of 550 nm irradiates the center of the grating, whose polarization direction is perpendicular to the grating grooves. The gold cone is embedded in air, and the numerical domain is surrounded by extra domains associated with the PML (IED) for electromagnetic (thermal) simulations. The same boundary conditions as those for the nanorod simulations are enforced. The height of the total numerical domain was 14 μm. The width and length were 7.5μm, as shown in Fig. S2(a). The layer thickness was 1 μm.

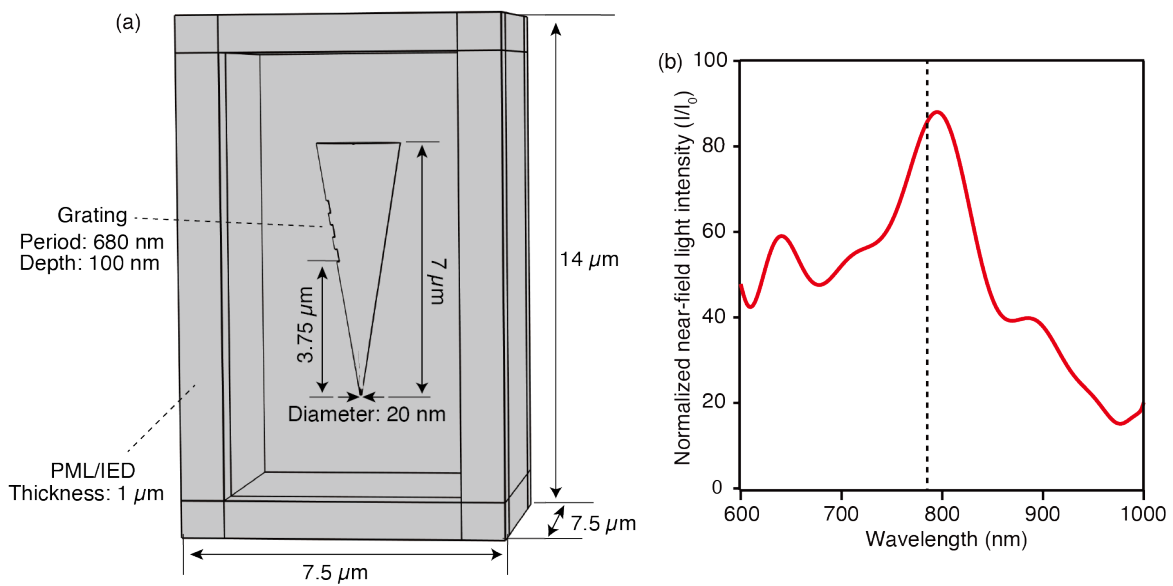

**Figure S2.** (a) Schematic of the calculation model for the gold cone. (b) Near-field light intensity spectrum of the gold cone normalized by incident light intensity.

# Note S4. Comparison between the temperature-dependent damping factor and the constant damping factor

As explained in Note S1, we considered the temperature-dependent damping factor to accurately represent realistic scenarios. To investigate the influence of the temperature-dependent damping factor, we compare our simulations with results obtained by considering a temperature-independent damping factor, and fixed to its value at room temperature. In the case where the damping factor is fixed, the permittivity is also constant at any temperature. As shown in Fig. S3, we calculated the near-field light intensities of the gold nanorod using temperature-dependent and temperature-independent damping factors. When the temperature-independent damping factor was considered, the near-field light intensity showed a linear change with respect to the incident light intensity, and reached much larger values than the case where we considered the damping factor varying with temperature. This clearly indicates that temperature significantly affects the damping factor, and that it is essential to consider the temperature-dependent damping factor, especially at a high temperature, that is, high incident light intensity.

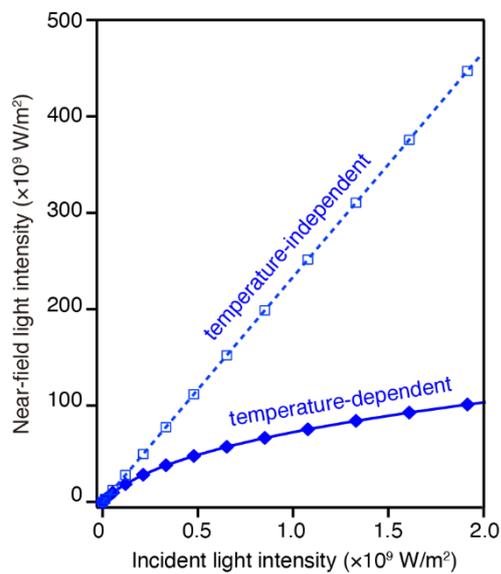

**Figure S3.** Relationships between the near-field light intensity and incident light intensity for the gold nanorod, calculated using temperature-dependent and temperature-independent damping factors.

## Note S5. The effect of the damping factor to the near-field light intensity

As shown in Fig. 3(i) in the main text, we observed an intensity decreases in the near-field light obtained by plasmon nanofocusing, even though the incident light intensity was increased. We ascribed this peculiar trend to an increase in the damping factor at high temperatures. To ascertain our interpretation, we carefully inspected the origin of the decrease in the near-field light intensity in the high illumination intensity regime. The intensity of near-field light is determined by two factors: (i) the intensity of the incident light; and (ii) the intensity enhancement from the incident light to the near-field light, through the process of plasmon resonance or nanofocusing. A higher incident light intensity simply provides a greater number of photons impinging on the gold nanorod or grating of the gold cone, which positively contributes to an increase in the near-field light intensity. Subsequently, however, a high incident light intensity increases the temperature and damping factor, which reduces the permittivity of gold ($\varepsilon$ = -24.123 + 1.552$i$). A smaller permittivity then contributes negatively to reducing the enhancement from the incident light intensity to the near-field light intensity. Therefore, the balance between these two factors determines the near-field light intensity.

Figure S4(a) shows the near-field intensity spectra generated by the localized plasmon resonance of the gold nanorod, normalized to the incident light intensity, which thus provides an enhancement of the near-field light intensity with respect to the incident light intensity. Here, the incident light intensities of 85.12 × $10^7$ and 224.77 × $10^7$ W/m$^2$ were used to excite the system, which led to an increased surface-averaged temperature of 895 and 1305 K, respectively. As a reference, the near-field intensity spectrum at room temperature is also shown as a blue curve. We can clearly see that the enhancement of the near-field light intensity drastically drops over the entire spectral range as the temperature increases. At a wavelength of 785 nm, the near-field light intensity is 59.38 times enhanced from the incident light intensity at 895 K (incident light intensity: 85.12 × $10^7$ W/m$^2$), whereas it is enhanced only by 37.03 times at 1305 K (incident light intensity: 224.77 × $10^7$ W/m$^2$). Therefore, while the incident light intensity is increased by ~2.64 times (~224.77 × $10^7$/85.12 × $10^7$), the enhancement decreases by ~1.60 times (~59.38/37.03). Overall, this results in an increase of the near-field light intensity by ~1.65 times.

In the case of plasmon nanofocusing, the normalized near-field light intensity spectra at temperatures of 895 and 1305 K are shown in Fig. S4(b), which required the incident light intensities of 505.73 × $10^7$ and 864.83 × $10^7$ W/m$^2$, respectively. Similarly, we found a drastic decrease in enhancement with temperature. At 895 K (required incident light intensity: 505.73 × $10^7$ W/m$^2$), the near-field light intensity is enhanced by 28.62 times, whereas it is enhanced only by 13.68 times at 1305 K (incident light intensity: 864.83 × $10^7$ W/m$^2$). Therefore, even though the incident light intensity is increased only by ~1.71 times (~864.83 × $10^7$/505.73 × $10^7$), the enhancement is decreased by ~2.09 times (~28.62/13.68). This results in a reduction of the near-field light intensity by 1.22 times although the incident light intensity increased. We confirmed that the large damping factor and the resulting smaller enhancement dominate plasmon nanofocusing to a larger extent than the case of localized plasmon resonance, which could lead to a weaker near-field light intensity at a higher incident light intensity.

In addition, we thoroughly investigated the trend of the near-field light intensity for more intense illumination. Figure S4(c) represents the ratio between the increment of the incident light intensity and the decrement of the near-field light enhancement, with respect to the incident light intensity. As such, if the ratio is greater than 1, the near-field light intensity increases as the incident light intensity increases. On the other hand, if it is less than 1, which means that the enhancement decrease is more dominant, the near-field light intensity decreases while the incident light intensity increases. As for the gold nanorod, it was calculated up to an incident light intensity of 2.60 × $10^9$ W/m$^2$, as the temperature goes beyond the melting point at an intensity greater than that. As the ratio was always greater than 1, the near-field light intensity monotonically increased with increasing incident light intensity, which agreed with the results shown in Fig. 2(g) in the main text. In contrast, in the case of plasmon nanofocusing of the gold cone, it is larger than 1 in the lower intensity range. However, it becomes less than 1 for an incident light intensity greater than 4.55 × $10^9$ W/m$^2$. This indicates that the near-field light intensity keeps decreasing with increasing incident light intensity for an incident light intensity higher than 4.55 × $10^9$ W/m$^2$, which is in good agreement with Fig. 3(i) in the main text. We quantitatively confirmed that in the case of plasmon nanofocusing, for an incident light intensity larger than a certain value, the near-field light intensity can be reduced even when the incident light intensity is increased.

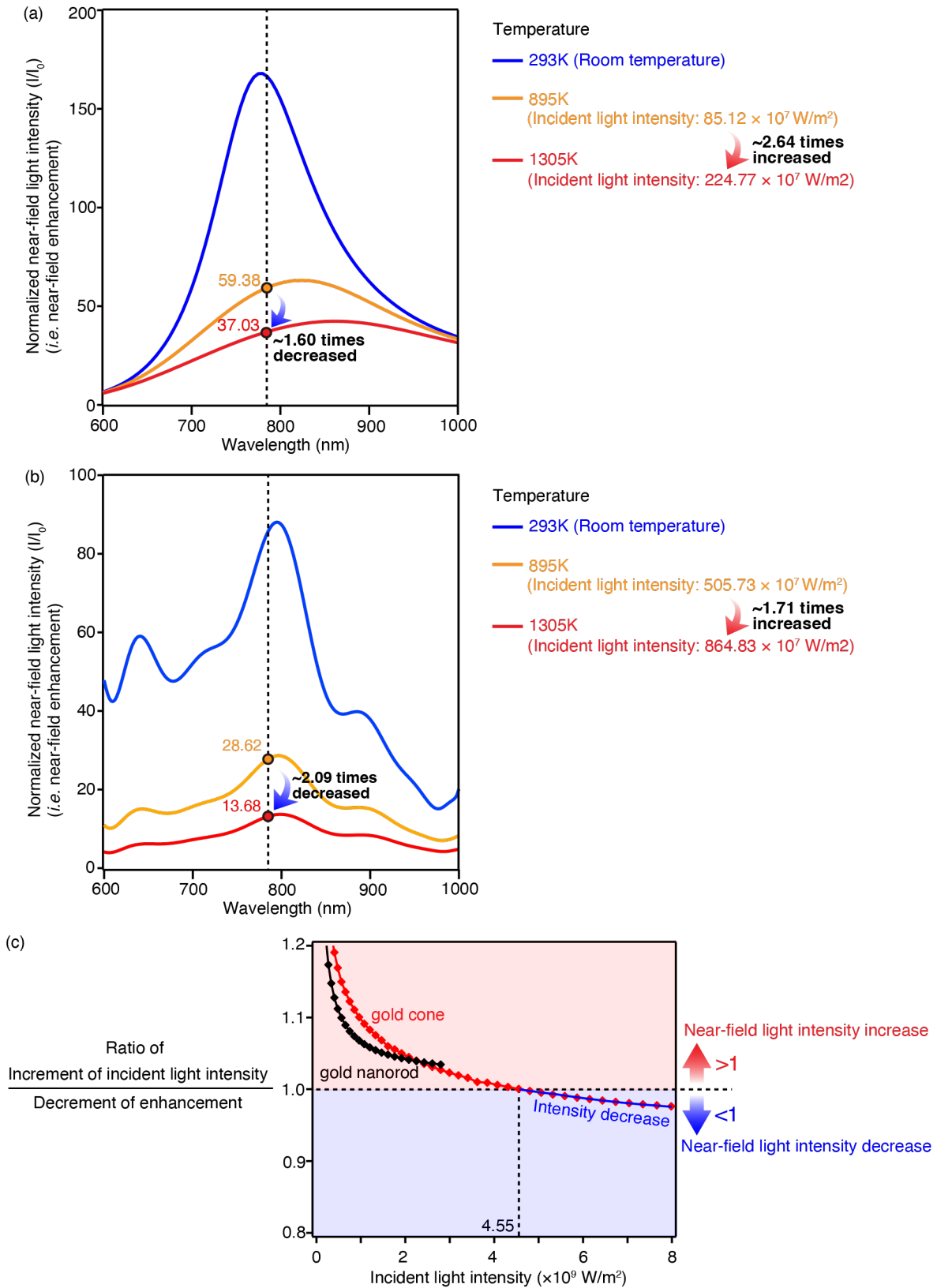

**Figure S4.** (a) Near-field light intensity spectra of the gold nanorod normalized by the incident light intensity at different temperatures. (b) Near-field light intensity spectra of the gold cone normalized by the incident light intensity at different temperatures. (c) Ratio of increment of incident light intensity to decrement of near-field light enhancement, plotted with respect to the incident light intensity.

**Reference**

[1] A. Alabastri, et al., *Materials* **6**, 4879-4910 (2013).